\documentclass[runningheads]{llncs}
\usepackage[utf8]{inputenc}
\usepackage[T1]{fontenc}
\usepackage{graphicx}
\usepackage{acronym}
\usepackage{ninecolors}
\usepackage{tikz}
\usepackage{tabularray}
\usepackage{listings}
\usetikzlibrary{shapes.geometric, arrows.meta, positioning, fit, backgrounds, calc}
\acrodef{RISE}[\texttt{RISE}]{Recursive Improvement via Self-play and Evolution}
\acrodef{DR}{Deep Research}
\acrodef{GEPA}{\textbf{Ge}netic-\textbf{Pa}reto}

\definecolor{pink}{HTML}{EB5CA1}
\definecolor{teal}{HTML}{469CA0}
\definecolor{zaorange}{HTML}{F28C28}
\newcommand{\orchestrator}{\texttt{orchestrator}}
\newcommand{\reader}{\texttt{reader}}
\newcommand{\aggregator}{\texttt{aggregator}}
\newcommand{\writer}{\texttt{writer}}

\DeclareRobustCommand{\circnum}[1]{%
  \tikz[baseline=(C.base)]\node[circle, fill=zaorange, inner sep=1.2pt, minimum size=1.8ex, text=white, font=\sffamily\bfseries\footnotesize] (C) {#1};%
}

\newcommand{\orchestratorc}{\textcolor{pink}{\orchestrator}}
\newcommand{\readerc}{\textcolor{teal}{\reader}}
\newcommand{\aggregatorc}{\textcolor{pink}{\aggregator}}
\newcommand{\writerc}{\textcolor{pink}{\writer}}

\begin{document}
\title{Self-Optimizing Multi-Agent Systems for Deep Research}
\author{Arthur C\^amara \and Vincent Slot \and Jakub Zavrel}
\institute{Zeta Alpha, Amsterdam, The Netherlands\\
  \email{\{camara,vincent.slot,zavrel\}@zeta-alpha.com}\\
\url{www.zeta-alpha.com}}
\maketitle

\begin{abstract}
  Given a user's complex information need, a multi-agent Deep Research system iteratively plans, retrieves, and synthesizes evidence across hundreds of documents to produce a high-quality answer. In one possible architecture, an \emph{orchestrator agent} coordinates the process, while parallel \emph{worker agents} execute tasks.
  Current Deep Research systems, however, often rely on hand-engineered prompts and static architectures, making improvement brittle, expensive, and time-consuming. We therefore explore various multi-agent optimization methods to show that enabling agents to self-play and explore different prompt combinations can produce high-quality Deep Research systems that match or outperform expert-crafted prompts.
\end{abstract}

\section{Introduction}\label{sec:intro}

When tackling complex information needs that require structured long-form answers, hundreds of documents and dozens of queries may be needed before enough information can be acquired to produce the final answer. Further complicating matters, this process is highly \textit{iterative}: the information gathered in one step informs subsequent queries and the research direction in later iterations. Such settings require retrieval systems to synthesize intermediate findings, identify knowledge gaps, and continuously formulate new queries. This paradigm, often referred to as \acf{DR}, moves away from the standard single-turn ``retrieve-then-generate'' workflow that is common in traditional Retrieval-Augmented Generation (RAG)~\cite{lewis2020rag}. Instead, \ac{DR} systems rely on an agent that can issue multiple queries, inspect large numbers of search results, and reason over the acquired knowledge before repeating this process until it can produce a high-quality answer~\cite{huang2025deep,zhang2024agentpro,shao2024storm}.

While reasonably successful, these systems remain heavily dependent on human intuition and heuristic design. Most current \ac{DR} systems rely on handcrafted architectures and verbose, static system prompts to guide agent behavior~\cite{stateact2025}. However, this approach is inherently brittle. 
A set of prompts that performs well for one Large Language Model (LLM) and one specific topic may fail when models change or when applied to a different field or research task altogether. When such changes inevitably occur, this typically triggers an expensive and time-consuming round of trial-and-error~\cite{spiess2025autopdl}, where developers manually tweak prompts and architectures until they reach a reasonable quality threshold~\cite{khattab2024dspy}.

Recently, it has been proposed to shift from manual ``prompt engineering'' towards a more algorithmic ``prompt optimization'' process. Such proposals---including GEPA~\cite{agrawal2025gepa}, TextGrad~\cite{yuksekgonul2024textgrad}, AutoPDL~\cite{spiess2025autopdl}, GASO~\cite{wang2024correctly}, GAPO~\cite{gu2025gapo}, the Darwin-Gödel Machine~\cite{zhang2025dgm}, and topology optimization methods~\cite{ramnath2025multiagent}---suggest that natural language prompts as well as other multi-agent system parameters and architectures can be treated as trainable parameters and optimized. By employing feedback mechanisms, where the frontier LLM itself is the optimization operator, these systems can self-improve: prompts can be refined, agents can explore and self-play, and high-quality agents can be produced with minimal manual input~\cite{dharna2025selfplay,hu2025automated,yang2024opro}.

In this work, we explore how these approaches, in particular TextGrad and GEPA, fare when applied to an industrial-scale \ac{DR} system. We next introduce and discuss our multi-agent architecture for \ac{DR} and show how a GEPA-inspired method can create high-quality agents that can outperform expert-crafted prompts.

\section{A Multi-Agent Deep Research Architecture}\label{sec:architecture}
\begin{figure*}[!t]
  \centering
  \includegraphics[width=\textwidth]{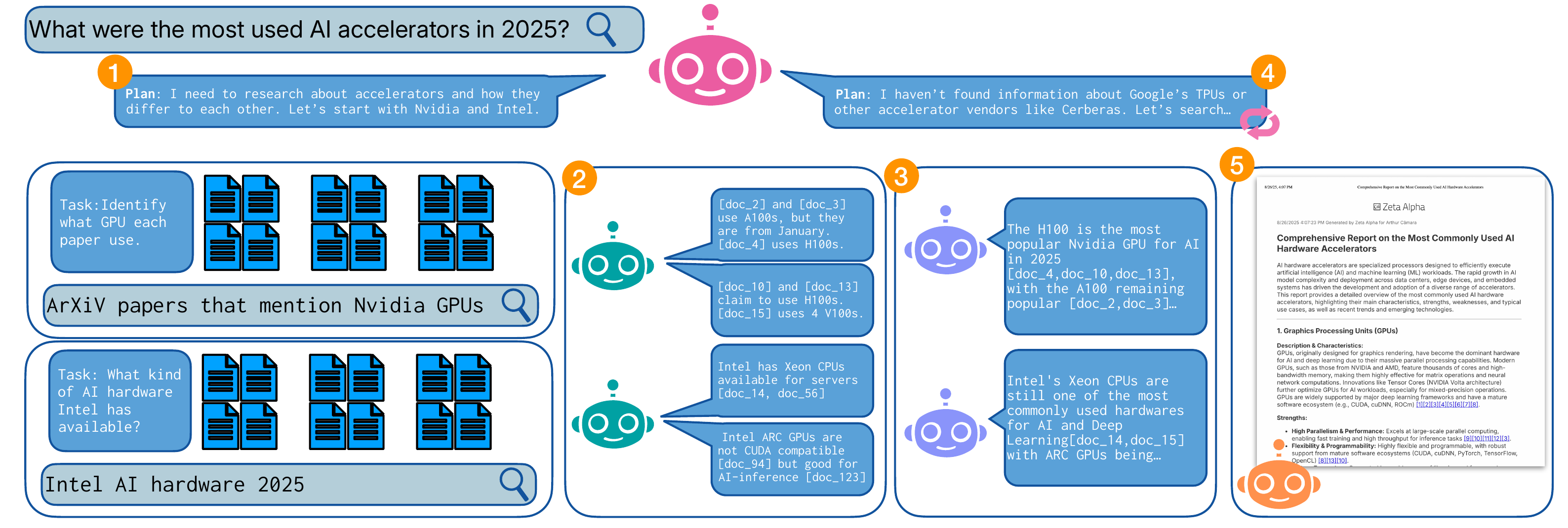}
  \caption{Architecture for a multi-agent \acl{DR} system: \circnum{1} an \emph{orchestrator agent} (\orchestratorc{}) creates a list of tasks for the user's question. Each task consists of a query and instructions. \circnum{2} multiple \emph{reader agents} (\readerc{}) inspect batches of documents and extract the information requested in the task. \circnum{3} an \emph{aggregator agent} (\aggregatorc{}) combines these smaller information pieces into larger mini-reports for each task. \circnum{4} the \emph{orchestrator agent} can decide whether to refine the plan with more tasks or \circnum{5} call a \emph{writer agent} (\writerc{}), which combines all merged information into a long-form report.}\label{fig:architecture}
\end{figure*}

In general, \acl{DR} systems contain four main components~\cite{shi2025survey}: (i) \emph{Query planning}: decomposing the initial complex research query formulated by the system's user into a series of simpler, manageable sub-queries; (ii) \emph{Information acquisition}: using external knowledge bases (e.g., a search engine) or other tools to gather relevant information as needed; (iii) \emph{Memory management}: maintaining and updating the context relevant to the task with the information already retrieved; and (iv) \emph{Answer generation}: producing comprehensive, source-attributed outputs, such as scientific reports or long-form answers.

Our instantiation of a multi-agent \ac{DR} system is shown in Figure~\ref{fig:architecture}. The numbered callouts in the figure correspond to the main control flow: \circnum{1} an \emph{orchestrator agent} (\orchestrator{}) produces an initial plan and a set of tasks (each with a search query and instructions); \circnum{2} multiple \emph{reader agents} (\reader{}) retrieve and read documents in parallel and extract task-relevant evidence snippets; \circnum{3} an \emph{aggregator agent} (\aggregator{}) consolidates these snippets into task-level mini-reports; \circnum{4} the \emph{orchestrator agent} updates its state (what is covered vs. what is missing) and either issues new tasks or proceeds to writing; and \circnum{5} a \emph{writer agent} (\writer{}) produces the final long-form report with citations.

In addition to these agents, our system includes orchestration and bookkeeping utilities (e.g., citation tracking, de-duplication, and cleaning) that support memory management and source attribution. We discuss our implementation in more detail in Section~\ref{sec:experiments}.\\

\noindent Our Multi-Agent \acl{DR} system contains the following agents:

\paragraph{Orchestrator.} Given the user query, the \orchestrator{} produces an initial plan for how to tackle the problem. It is composed of a natural language description of the plan and a list of \emph{tasks} that the reader agents should accomplish. Each task contains (a) a search query (formulated as a tool call) and (b) a concise description of what type of information the reader agents should look for when inspecting retrieved documents. After each round of exploration and consolidation, the orchestrator agent analyzes the information collected so far and either refines the plan by issuing new tasks or stops and proceeds to the \writer{}.

\paragraph{Reader.} For each task, the system retrieves a large number of documents (e.g., webpages, PDFs, or internal knowledge base entries or documents in an Enterprise Search setting). Multiple parallel \reader{} agents then analyze these documents in batches, extracting task-relevant information according to the instructions defined by the \orchestrator{}. Each \reader{} agent outputs citation-backed evidence snippets that can be traced back to the source documents (for later citation and de-duplication handling). This stage is designed for high recall, allowing hundreds of documents to be inspected in parallel.

\paragraph{Aggregator.} The \aggregator{} agent combines the information extracted by all \reader{} agents for a given task, producing a \emph{mini-report} with the information requested by the \orchestrator{} for that task. The intuition is that the \aggregator{} agent can de-duplicate overlapping evidence extracted by multiple \reader{}s, reconcile or surface conflicts, and discard weak evidence.

\paragraph{Writer.} At some point, the \orchestrator{} agent either decides that it has gathered enough information to produce a final report, or reaches a pre-defined limit of compute resources, and invokes the \writer{} to generate the long-form report by combining the information compiled in the mini-reports produced by the \aggregator{} agents. Here, our system also handles citation insertion and produces a final ``References'' section.

\paragraph{Citation management.} Throughout the pipeline, each piece of extracted evidence is tagged with a unique document identifier (i.e., a hash of its URL) linking it back to its source. The \reader{} agents output citation-backed snippets using a consistent format (e.g., square-bracketed identifiers). The \aggregator{} preserves these identifiers when merging information and de-duplicates overlapping evidence from the same source. Finally, the \writer{} resolves all identifiers into a unified bibliography, ensuring that each claim in the final report can be traced to its provenance.
\section{Self-optimizing Deep Research systems}\label{sec:optimization}
As discussed in Section~\ref{sec:intro}, each agent in a \acf{DR} system typically contains a hard-coded prompt written by a human expert and tuned to a specific setting and domain. While this works well when the system is deployed in similar conditions, it makes the agents brittle to changes in the domain or model. For example, a \ac{DR} system designed for chemical engineering may contain domain-specific terms and instructions that do not generalize to a biomedical setting. Similarly, changing the underlying model (e.g., from a closed-source LLM to a locally deployed LLM) can cause significant variation in the quality of the results produced by the system.

Such changes lead to a long, difficult, and costly process of adapting individual prompts to a new domain, model, or shifts in user behavior. The alternative we propose in this work, is to \emph{automatically optimize} the multi-agents system, allowing the system to self-play and explore variations and combinations of agent prompts with minimal manual intervention. In this paper, we focus optimization of prompts, although all aspects of the multi-agent system (parameters, LLM models, system archtecture, etc.) can be optimized using the same principles.

More specifically, we optimize a candidate system $C=\{c_o, c_r, c_a, c_w\}$ using a training set $\mathcal{D}=\{(q_i,\mathcal{R}_i)\}_{i=1}^N$, where $\mathcal{R}_i=\{r_{i,1},\dots,r_{i,K_i}\}$. Here, $c_o, c_r, c_a,$ and $c_w$ are the system prompts for the \orchestrator{}, \reader{}, \aggregator{}, and \writer{}, respectively; $q_i$ is a training query; and $\mathcal{R}_i$ is the set of rubric criteria used to evaluate candidate answers for $q_i$.

In each optimization step $j$, we select a system $C$ from a pool of candidates and optimize one of the agents in $C$ (e.g., $c_o$), selected in a round-robin fashion. We then sample a mini-batch $\mathcal{B}\subset \mathcal{D}$ and generate answers for its queries using $C$. The optimization process outputs an updated prompt for the selected agent, yielding a new candidate system $C'_j$ (e.g., $\{c'_o, c_r, c_a, c_w\}$).

We explore two optimization methods applied to the \ac{DR} architecture discussed in Section~\ref{sec:architecture}: \ac{GEPA}~\cite{agrawal2025gepa} and TextGrad~\cite{yuksekgonul2024textgrad}. While the two methods differ in their details, both follow a similar \emph{reflection-to-refinement} pattern: the system's outputs (and, optionally, execution traces) are fed into a ``meta-prompt'' that produces suggestions for improving the prompt of a chosen agent. The main differences lie in how they select a candidate system $C$ from the pool and in the update procedure.

TextGrad~\cite{yuksekgonul2024textgrad} uses a greedy strategy inspired by the hill-climbing operator of backpropagation when selecting $C$. At each step, it selects the best current system as evaluated on a development split. When optimizing, it relies on a numerical gradient-descent metaphor: given a candidate $C$, the selected agent prompt is treated as a ``learnable parameter'' and updated via ``textual gradient descent.'' In practice, for each sample in $\mathcal{B}$, the final answer is evaluated against its rubric. The evaluation signal and the agent execution trace are then fed into an LLM that produces a ``loss'' (i.e., a critique of the prompt based on the outcome). In a backward pass, the losses across $\mathcal{B}$ are aggregated (e.g., concatenated) to form a ``gradient'' (i.e., critiques of how the prompt could be improved). Finally, an optimizer updates the prompt based on these gradients.

For \ac{GEPA}~\cite{agrawal2025gepa}, the candidate system $C$ is selected using a Pareto-based procedure by a genetic algorithm from a population of reasonably good solutions. Given the training queries, \ac{GEPA} prunes strictly dominated candidates (i.e., candidates that are not the best on any query) and samples candidates with probability proportional to their Pareto ``support'' (i.e., the number of instances in which they appear on the frontier). We then select one agent to optimize and generate agent-specific feedback for each query in $\mathcal{B}$. This feedback is used to generate a new prompt for the selected agent, yielding a new candidate system $C'$. Another key difference from TextGrad is that $C'$ is discarded if it does not improve over $C$ on $\mathcal{B}$.

\section{Experiments}\label{sec:experiments}

For our experiments, we adapt the experimental setup proposed by ScholarQA~\cite{asai2024openscholar}. Specifically, we use the Computer Science subset of queries. This dataset comprises 109 query--rubric pairs written by PhD-level experts in Computer Science. For each query, the rubrics contain a list of criteria that constitute a ``good'' answer, marked as either ``most important'' or ``nice to have'' items. The evaluation is performed by an LLM-as-judge on a per-rubric basis. We split the dataset into 29 pairs for training, 50 for testing, and 30 for development. The results are shown in Table~\ref{tab:results}.

\begin{table}[h]
  \centering
  \caption{Average scores on ScholarQA-CS for \acl{DR} systems optimized using TextGrad, \ac{GEPA} with the default meta-prompt, \ac{GEPA} with a custom meta-prompt, or OpenAI's prompt optimizer for GPT-4.1.}\label{tab:results}
  \begin{tblr}{
      width=\linewidth,
      colspec={X[c] X[c] X[c] X[c] X[c]},
      colsep=1pt,
      row{odd}={bg=azure9},
      row{1} = {c, bg=azure3, fg=white, font=\sffamily, cmd=\textbf},
      column{1} = {c, bg=azure3, fg=white, font=\sffamily, cmd=\textbf\small},
      hline{1,Z} = {solid, 1pt},
      hline{3,4} = {2-2}{solid, 0.5pt},
      cell{2}{2} = {r=4, c=1}{c},
      cell{2}{4} = {r=4, c=1}{c},
    }
    Optimizer & Minimal Prompt  & Minimal Prompt + Optimizer    & Expert Prompt & Expert Prompt + Optimizer \\
    TextGrad            & 0.513 & 0.654             & 0.667 & 0.672             \\
    \ac{GEPA}           &       & 0.685             &       & 0.670    \\
    \ac{GEPA} custom    &       & \textbf{0.705}    &       & \textbf{0.701}    \\
    OpenAI              &       & 0.583             &       & 0.667             \\
  \end{tblr}%
\end{table}

We compare four different optimizers: TextGrad, \ac{GEPA} with the default meta-prompt as stated in the original paper, \ac{GEPA} with a custom meta-prompt for the \ac{DR} task, and OpenAI's prompt optimizer~\footnote{\url{https://platform.openai.com/chat/edit?models=gpt-4.1\&optimize=true}}. The agents are implemented using Zeta Alpha's Agents SDK~\footnote{\url{https://github.com/zetaalphavector/platform/tree/master/agents-sdk}}. We use TextGrad's library to implement its optimizer~\footnote{\url{https://github.com/zou-group/textgrad}}. For \ac{GEPA}, an open-source implementation was not available when this project started, so we implemented it ourselves. For search, we use the Deep Research Gym API with the FineWeb index~\cite{coelho2025deepresearchgym}. All agents, optimizers, and judges are instantiated with GPT-4.1-mini. The custom meta-prompt for \ac{GEPA} is included in Appendix~\ref{sec:meta_prompts}.

We initialize our agents with one of two settings: Either a minimal, one-liner prompt or a prompt written and refined by experts for more than one year (i.e., the prompt used in Zeta Alpha's Deep Research agent at the time). The minimal prompts are included in Appendix~\ref{sec:prompts}.

\paragraph{Hyperparameters.} All experiments share the same hyperparameters. All agents, judges, and optimizers use GPT-4.1-mini with a mini-batch size of 3. We also experimented with GPT-5 and Qwen3 models; however, their shorter context windows (400,000 and 32,768 tokens, respectively, versus 1,047,576 for GPT-4.1-mini) required drastically reducing trace lengths and batch sizes, which led to poorer initial results and other issues. All optimizers had the same maximum budget of USD\,50.00 per round. For TextGrad, we ran optimization for up to 10 epochs. For \ac{GEPA}, we set a patience limit of 2 (i.e., the number of consecutive rounds with no improvement on the development set before stopping). 

\paragraph{Evaluation protocol.} Following ScholarQA, we employ an LLM-as-judge approach. Each rubric item defines a criterion (e.g., ``covers key algorithmic trade-offs'') and, optionally, supporting evidence snippets that a good answer should incorporate. The judge scores each criterion on a 0--10 scale, normalized to $[0,1]$. In addition to the rubric-specific criteria (60\% of the final score), we evaluate three static components: (i)~\emph{expertise alignment} (10\%)---whether the answer's complexity matches the expected audience; (ii)~\emph{citation coverage} (20\%)---the fraction of claims backed by a citation; and (iii)~\emph{excerpt presence} (10\%)---the fraction of citations accompanied by verbatim excerpts. The weighted sum of these components yields the final score reported in Table~\ref{tab:results}.

\paragraph{Results.} Table~\ref{tab:results} shows that prompt optimization consistently improves performance relative to the minimal prompt baseline. With TextGrad, adding the optimizer increases the minimal-prompt score from 0.513 to 0.654 (+0.141), approaching the expert-prompt setting (0.667). Notably, the gains from optimization are smaller once a strong prompt is already in place: optimizing the expert prompt yields a modest improvement (0.667 to 0.672, +0.005). This suggests that, at least for this setup, optimization is most valuable when starting from weaker initial prompts. 

We also observed that the evolutionary aspect of \ac{GEPA}, combined with Pareto-optimal exploration, leads to more efficient search, as illustrated by the exploration tree in Appendix~\ref{sec:exploration}. This results in a more efficient exploration of the prompt space and converges to a strong system more quickly than TextGrad's greedy search.

Among the optimizers, \ac{GEPA} performs strongly across both prompt regimes. Using \ac{GEPA} with the default meta-prompt yields 0.685 (minimal prompt + optimizer) and 0.670 (expert prompt + optimizer), outperforming TextGrad in the minimal-prompt regime and matching it in the expert-prompt regime. The best results are obtained with \ac{GEPA} using a task-specific meta-prompt, reaching 0.705 and 0.701 for the minimal and expert prompts, respectively. Compared to the strongest non-\ac{GEPA} setting (TextGrad with expert prompt + optimizer at 0.672), this represents a sizable margin, indicating that tailoring the meta-prompt to the Deep Research task can translate into measurable end-to-end improvements.

Interestingly, OpenAI's built-in prompt optimizer underperforms relative to both TextGrad and \ac{GEPA}. Starting from the minimal prompt, it achieves only 0.583---substantially below TextGrad's 0.654 and \ac{GEPA}'s 0.685. When initialized with the expert prompt, OpenAI's optimizer shows no improvement (0.667), merely matching the unoptimized baseline. We hypothesize that this gap stems from the optimizer's design: OpenAI's tool is a general-purpose prompt improver without access to task-specific evaluation signals or execution traces. In contrast, both TextGrad and \ac{GEPA} leverage rubric-based feedback and agent traces to guide refinement, enabling more targeted improvements for the \ac{DR} setting.

\section{Discussion and future work}\label{sec:discussion}

Our experiments suggest that prompt optimization can be an effective mechanism for improving \acl{DR}-like Multi-Agent Systems, particularly when starting from minimal prompts. In this regime, both TextGrad and \ac{GEPA} substantially improve performance over the unoptimized baseline, matching or even surpassing expert-generated prompts. This indicates that optimization can be a viable path to creating high-quality multi-agent \ac{DR} systems with minimal manual intervention. At the same time, the gains from optimizing already strong expert prompts are smaller, suggesting diminishing returns once a prompt captures the key task structure.

A second takeaway is that the optimization procedure itself matters. In our results, \ac{GEPA} outperforms TextGrad, and tailoring the meta-prompt to the \acl{DR} setting yields the best overall performance. This highlights that ``prompt optimization'' is not a plug-and-play replacement for expert design: the choice of objective, selection strategy, and meta-prompt can significantly affect the quality and stability of the resulting agents.

\paragraph{Limitations.} Our study has several limitations. First, we evaluate on a single domain (Computer Science) with a relatively small dataset (109 queries, 50 for testing), which limits generalizability claims. Second, we rely on a single model family (GPT-4.1-mini) for agents, optimizers, and judges; different model combinations may yield different conclusions. Third, LLM-as-judge evaluation, while scalable, may introduce biases (e.g., favoring verbose or stylistically similar outputs). Finally, we do not report statistical significance tests; given the small test set, some differences may not be robust.

\paragraph{Future work.} While promising, this work remains exploratory. We see three clear directions. First, expand \emph{what} gets optimized beyond agent prompts, e.g., allowing the optimizer to select tools, adjust architectural hyperparameters, or even propose code-level changes, in the spirit of systems such as DGM~\cite{zhang2025dgm}. Second, reduce reliance on expert supervision by moving toward synthetic or self-generated training signals; recent work such as Dr.~Tulu~\cite{shao2025drtulu} and ResearchRubrics~\cite{sharma2025researchrubrics} suggests this can be feasible. In this setting, alternative candidate-selection strategies (e.g., long-running Elo-style tournaments~\cite{rackauckas202ragelo}) may further improve exploration. Third, extend these optimization frameworks beyond \ac{DR} to other agentic tasks, particularly those that can be evaluated with scalable, synthetic query-and-evaluation pipelines.

\bibliographystyle{splncs04}
\bibliography{references}
\appendix
\section{Minimal prompts}\label{sec:prompts}
Minimal \orchestrator{} prompt:
\begin{lstlisting}
Given a user query, create a report that answer the user's question.
\end{lstlisting}

Minimal \reader{} prompt:
\begin{lstlisting}
Given a user's question, a search query, a list of search results and instructions on what information to retrieve, extract any information that is relevant to the user's question.

For any result that you use in your response, you must cite it using square brackets with the provided document_id.
\end{lstlisting}
Minimal \aggregator{} prompt:
\begin{lstlisting}
Given a user question, a set of instructions on what information to extract and the list of extracted information pieces with supporting evidence, combine the information into a single answer that fulfills the instructions.
\end{lstlisting}

Minimal \writer{} prompt:
\begin{lstlisting}
Given a user question and a list of information pieces, write a report that answers the user's question.\end{lstlisting}
\section{Meta prompt}\label{sec:meta_prompts}
For the \ac{GEPA} optimizer with a custom meta-prompt, we used the following system prompt:
\begin{lstlisting}
  Your task is to optimize and refine individual agents of a Deep Research system. The primary goal of this system is to provide high-quality answers to user's questions by searching and combining information found in multiple documents.
  The system has the following agents: `orchestrator`, `reader`, `aggregator` and `writer`.

  The workflow of the system is the following:
    1. The `orchestrator` receives a user's question and devises a plan with a list of research tasks that need to be completed before writing the final report.
    2. Each task's query is submitted to a search engine, and the relevant information from each results page is extracted by the `reader`. The information of all search results pages is then combined by the `aggregator`.
    3. The `orchestrator` reads the merged information for all submitted tasks and decides to either run another round of tasks or call the `writer`.
    4. The `writer` receives all the information from the tasks and writes a final report.

  I will provide you with a list of examples of different task inputs provided to a single agent, together with some feedback on the quality of the output generated by the agent using its current instructions.

  Read the inputs carefully and identify the input format and infer detailed task description about the task I wish to solve with the system.

  Read all the agent responses and the corresponding feedback. Identify all niche and domain specific factual information about the task and include it in the instruction, as a lot of it may not be available to the agent in the future. The agent may have utilized a generalizable strategy to solve the task, if so, include that in the instruction as well.

  Rules:
    1. Make sure your new instructions are generalizable to any computer science related task, and not specific to any particular task present in the examples.
    2. Do not include any other information or comments in your response.
    3. Do not suggest or imply any formatting to the output of the agent, like requiring the output to be a JSON or have specific fields, unless this is already present in the current instructions.

  In this round, you are optimizing the prompt of the `{{ agent_name }}` agent.
\end{lstlisting}


\section{Exploration trees}\label{sec:exploration}
\begin{figure*}[h!]
  \centering
  \includegraphics[width=\textwidth]{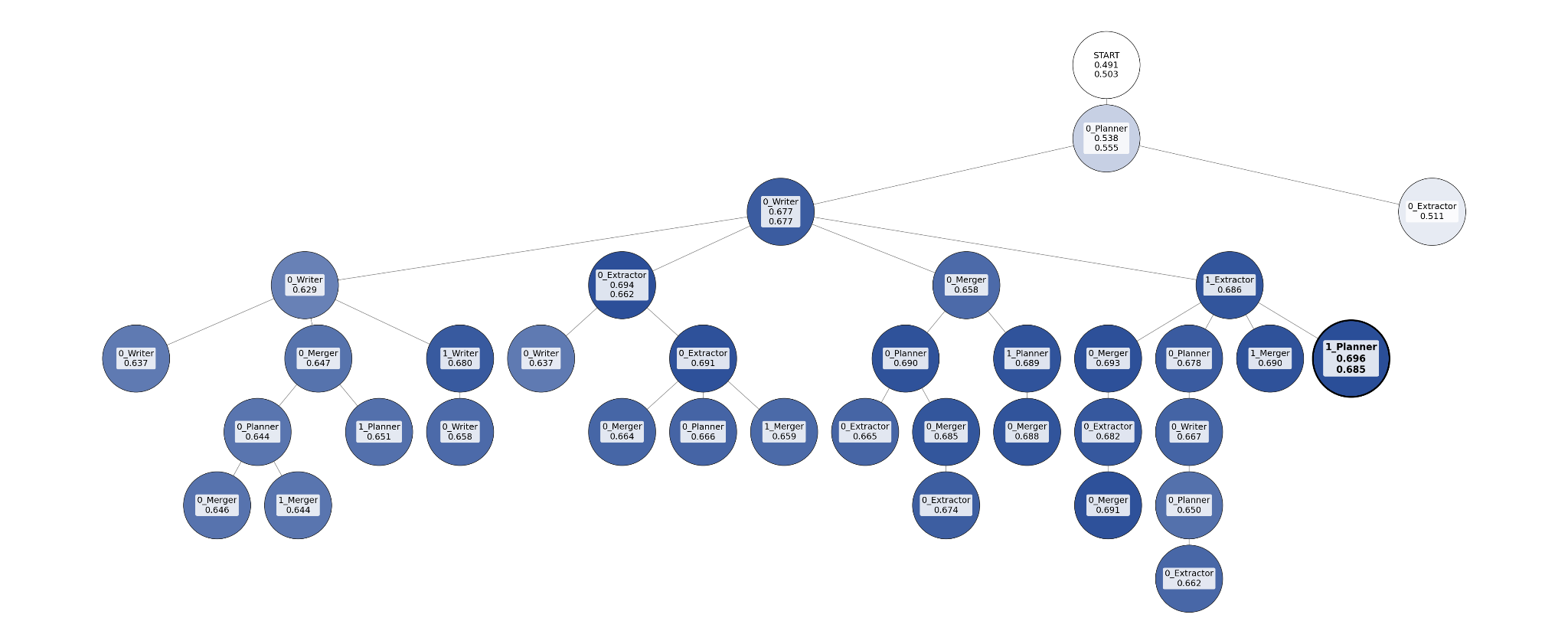}

  \vspace{1em}
  \includegraphics[width=\textwidth]{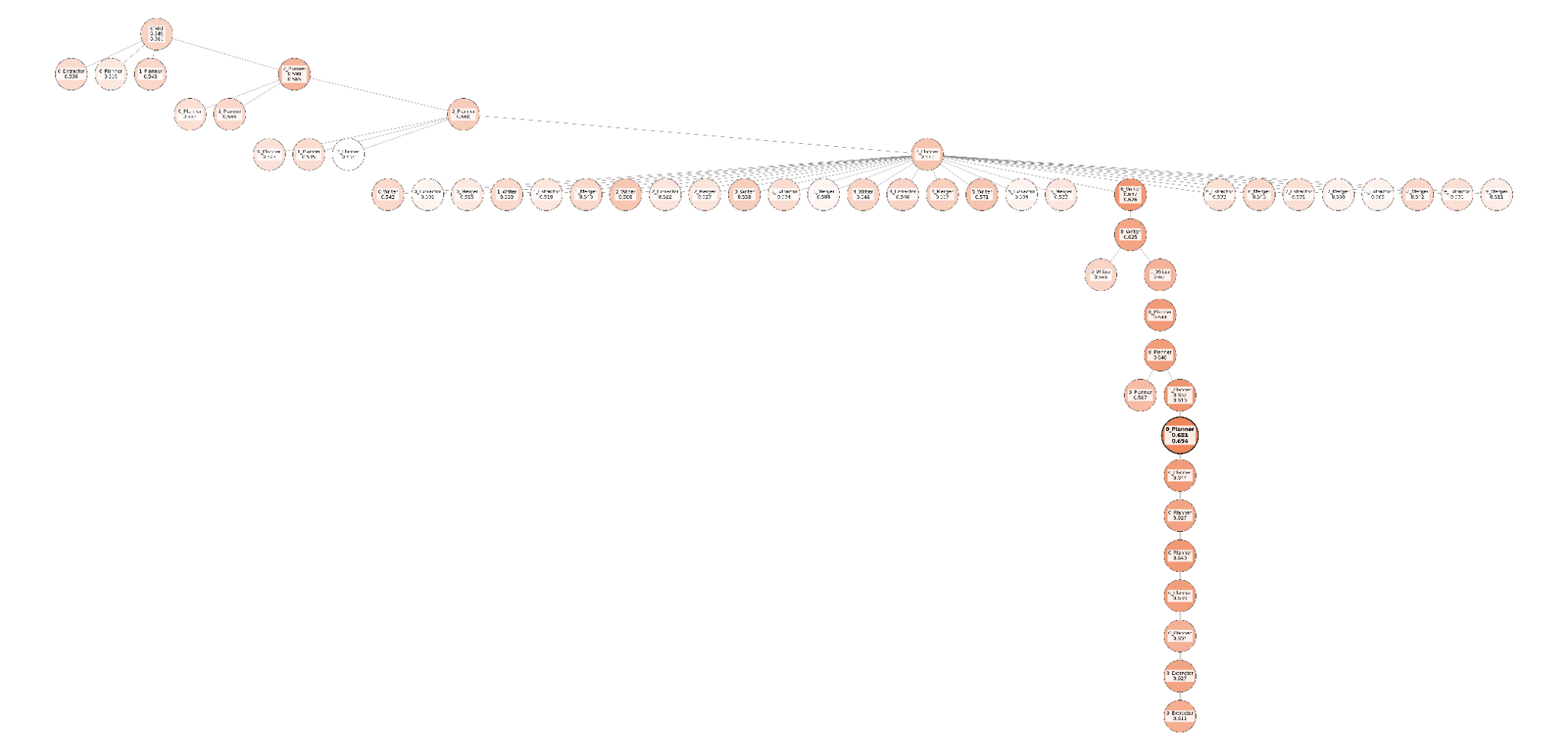}
  \caption{Example of exploration trees for both \ac{GEPA} and TextGrad. Each node in the tree is a new candidate that was generated based on its parent. \ac{GEPA} manages to explore different variants in a more diversified manner, while TextGrad does not explore that much.}\label{fig:trees}

\end{figure*}

Figure~\ref{fig:trees} shows an example of how \ac{GEPA} and TextGrad differ in their exploration of the prompt space.

\end{document}